\documentclass[twocolumn,showpacs]{revtex4}

\usepackage{graphicx}%
\usepackage{dcolumn}

\newcommand{\bb}{\begin{equation}}
\newcommand{\en}{\end{equation}}
\begin{document}

\title{Counterion Condensation and Fluctuation-Induced Attraction}

\author{A.W.C. Lau$^{1}$}
\author{P. Pincus$^{2}$}
\affiliation{$^{1}$Department of Physics and Astronomy, University of Pennsylvania, Philadelphia, PA 19104\\
$^2$Materials Research Laboratory, University of California, Santa
Barbara, CA 93106--9530}

\date{\today}

\begin{abstract}
We consider an overall neutral system consisting of two similarly
charged plates and their oppositely charged counterions and analyze the
electrostatic interaction between the two surfaces beyond the
mean-field Poisson-Boltzmann approximation.  Our physical picture is
based on the fluctuation-driven counterion condensation model, in which a fraction
of the counterions is allowed to ``condense'' onto the charged plates.  In
addition, an expression for the pressure is derived, which includes
fluctuation contributions of the whole system.  We find that
for sufficiently high surface charges,
the distance at which the attraction, arising from charge fluctuations,
starts to dominate can be large compared to the Gouy-Chapmann length.
We also demonstrate that depending on the valency, the system may exhibit a novel
first-order binding transition at short distances.
\end{abstract}

\maketitle

\section{Introduction}
\label{intro}

Correlation effects may play an important role in controlling the
structure and phase behavior of highly charged macroions in solutions
\cite{nature}.  The macroions may be charged membranes, stiff polyelectrolytes such as
DNA, or charged colloidal particles.  Recently, these effects have attracted
a great deal of attention, since they may drastically alter the standard
mean-field Poisson-Boltzmann (PB) picture \cite{rb,Ha,fluct,netz,shklovskii}.
For example, one surprising phenomenon is the {\em attraction}
between two highly-charged macroions, as observed in experiments \cite{attractionE,long,grier}
and in simulations \cite{Guldbrand,attractionS,messina}.
This attraction is not contained in the mean-field (PB) theory, even
for an idealized system of two charged planar surfaces.  Indeed,
it has been proven that PB theory predicts only repulsion between
likely-charged macroions \cite{neu}.

Very recently, another interesting effect that is not captured within
the PB theory is predicted, namely the novel fluctuation-driven
counterion condensation \cite{cond}.
For a system consisting of a single charged surface and its oppositely
charged counterions, Netz and Orland \cite{netz} showed that a simple
perturbative expansion about the mean-field PB solution breaks down for
sufficiently high surface charge. Thus, in this limit, fluctuation and
correlation corrections can become so large that the
solution to the PB equation is no longer a good approximation.
To circumvent this difficulty, a {\em two-fluid} model was proposed in
Ref. \cite{cond}, in which the counterions are divided into a {\em free}
and a {\em condensed} fraction.  The free counterions have the usual
3D mean field spatial distribution, while the
{\em condensed} counterions are confined to move only on the charged
surface and thus effectively reduce its surface charge density.  The number of
condensed counterions is determined self-consistently, by minimizing
the total free energy which includes {\em fluctuation} contributions.
This theory predicts that if surface charge density of the plate is
sufficiently high, a large fraction of counterions is ``condensed'' via
a phase transition, similar to the liquid-gas transition with a line of
first-order phase transitions terminating
at the critical point.  Furthermore, the valence of the
counterions plays a crucial role in determining the nature of the
condensation transition.  The physical mechanism leading to this
counterion condensation is the additional binding arising from 2D
charge-fluctuations, which dominate the system at high surface charge.
In  this paper, we extend this condensation picture to
a system of two charged surfaces with their neutralizing counterions,
and to study the electrostatic interaction between them.

Previous theoretical approaches to the problem of the attraction in
charged surfaces includes both numerical and analytical methods that go beyond
the mean-field PB theory.  Gulbrand {\em et al.} \cite{Guldbrand}
provides the first convincing demonstration for the attraction between
highly-charged walls using Monte Carlo simulations.  In particular,
they showed that for {\em divalent} counterions, the pressure between
charged walls becomes negative for distances less
than $10$ \AA;  Hence, the existence of short-ranged attraction.
Subsequently, there has been a number of numerical studies based
on the hypernetted chain integral equation \cite{integral} and the
local density-functional theory \cite{density}, as well as analytic
perturbative expansion around the PB solution \cite{ass} that
demonstrate attraction.

More recently, motivated by the problems of DNA condensation and
membrane adhesion,  two distinct approaches have been proposed to
account for the attraction arising from correlations \cite{rb,Ha,fluct}.
The first approach based on ``structural'' correlations first proposed
by Rouzina and Bloomfield \cite{rb}, the attraction comes from the
ground state configuration of the ``condensed'' counterions.
This theory predicts a strong short-ranged attraction, with the
characteristic length set by the lattice constant, typically of the
order of few Angstroms.  In the other approach, based on
charge-fluctuations, the counterion fluctuations are approximated by
the 2D Debye-H\"{u}ckel theory, which predicts a long-ranged attraction which
scales with the interplanar distance as $d^{-3}$.  Note, however, that the
mean-field PB repulsion which scales like $d^{-2}$ always dominates
the attraction for large distances, and thus, the range of the attraction
in this picture is still short, typically of the order of $10$ \AA  \cite{fluct2,zhou}.
Despite of the fact that some conceptual issues have been resolved concerning
the crossover of the attractions from long-ranged to short-ranged \cite{lau, Ha2}, there
remain some interesting problems to be understood.  In particular, some experimental
observations in planar surfaces \cite{long} and in charged colloidal suspensions \cite{grier}
as well as computer simulations \cite{messina} provide evidence for a
long-ranged attraction, typically of the order of microns, whereas the
two mechanisms mentioned above give only short-ranged attraction.
In this paper, we show that the charge-fluctuation approach,
together with the counterion condensation mechanism \cite{cond}, can
induce long-ranged attractions for sufficiently high surface charge.
We note that other mechanisms based on hydrodynamic interactions \cite{brenner},
depletion effects \cite{mario}, and an exact calculation for the 2D plasma model \cite{steve}
have been proposed recently to account for the long-ranged attractions.

In particular, we study the interaction between two charged surfaces
separated by a distance $d$, with counterions distributed both inside
and outside of the gap.  This boundary condition, as opposed to all of
the counterions confined between the gap, is more appropriate in general,
since systems are not closed and often the counterions are in equilibrium
with a ``bath'' in surface forces experiments.   In the spirit of
the ``two-fluid'' model proposed in Ref. \cite{cond}, we divide
the counterions into a ``condensed'' and a ``free'' fraction.
The condensed counterions are allowed to move only on the charged
surfaces, while the free counterions distribute in the space inside
and outside the gap.  The surface density of the condensed counterion
$n_c$ on each plate is determined by minimizing the total free energy,
which includes fluctuation contributions.
Furthermore, an expression for the fluctuation pressure is
derived, which includes fluctuation contributions from the condensed and ``free''
counterions, and their couplings.  We find that the counterion condensation can
occur either by increasing surface charge density at a fixed distance
or by decreasing the separation between plates.  For low surface
charge, the counterion condensation proceeds continuously as a function
of distance with the fraction of counterion condensed being small but finite, and
the total pressure of the system remains repulsive.

For higher surface charges, the qualitative behavior of the counterion
condensation transition depends critically on the valence $Z$ of the
counterions. For $Z < 2$, the counterion condensation proceeds continuously
as a function of distance.  On the other hand, for $Z \geq 2$,
the behavior of the system is qualitatively different, similar to
an isolated charged plate \cite{cond}.  In this case, the counterion
condensation occurs via a first-order phase transition as a function of distance.
Remarkably, we find that for {\em trivalent} ($Z=3$) counterions,
there is a wide range in the surface density, in which the first-order
counterion condensation spontaneously takes that the system from a repulsive
regime to an attractive regime at short distances, resulting in a first-order
binding transition.  For high surface charge, counterion condensation again proceeds
continuously even for $Z \geq 2$, but with a significant number of
condensed counterions.  Thus, in this regime, the mean-field repulsion is
substantially reduced and the long-ranged charge-fluctuation attraction dominates
the system even for large distances.
Note, however, that the mean-field repulsion will eventually
dominate as $d \rightarrow \infty$.  We emphasize that all these
features, in particular the special role of the valence,
deviate significantly from the PB mean-field predictions.

This paper is organized as follows: In Sec. \ref{sec:condensation}, we
briefly recapitulate qualitatively the mechanism which drives the counterion
condensation. In Sec. \ref{sec:twofluid}, we present in detail the two-fluid model
and derive a general expression for the total free energy of the counterions.  In Sec.
\ref{sec:twoplate}, we apply this formalism to study the interaction of similarly charged
surfaces. A detailed discussion of our results is presented in Sec. \ref{sec:results}.

\section{Counterion Condensation: Qualitative Argument}
\label{sec:condensation}

In this section, we recapitulate the essential physics of
the condensation transition presented in Ref. \cite{cond}.
Recall that for a single plate of charge density
$\sigma({\bf x}) = e n_0 \delta(z)$ immersed in an aqueous solution
of dielectric constant $\epsilon$, containing oppositely charged $- Z e$
point-like counterions of valence $Z$ on both sides of the plate,
PB theory predicts that the counterion distribution \cite{sam}\bb
\rho_0(z) =  {1   \over 2 \pi Z^2 l_B\left ( |z| + \lambda \right )^2},
\label{scd}
\en
decays to zero algebraically with a characteristic length
$\lambda \equiv  { 1 /(\pi l_B Z n_0)}$, where
$l_B \equiv {e^{2} \over \epsilon k_{B}T} \approx 7\,$\AA$\,\,$
is the Bjerrum length in water at room temperature,
$k_B$ is the Boltzmann constant, and $T$ is the temperature.
This Gouy-Chapman length $\lambda$ defines a sheath near the
charged surface within which most of the counterions are confined.
Typically, it is on the order of few angstroms for a moderate charge density of
$n_0 \sim 1/100\,\,$\AA$^{-2}$. Note that since
$\lambda$ scales inversely with $n_0$ and linearly with $T$,
at sufficiently high density or low temperature, the counterion
distribution is essentially two-dimensional. In fact, in the limit $T
\rightarrow 0$, we have\bb
\lim_{T \rightarrow 0}\,\int_{-\zeta}^{\zeta}\,\rho_0(z)\,dz\, =
\lim_{T \rightarrow 0}\,2\cdot{n_0 \over 2 Z}\cdot{ \zeta
\over \lambda + \zeta} = {n_0 \over  Z}, \nonumber
\en
where $\zeta$ is an arbitrarily small but fixed positive value of
$z$, {\em i.e.} the counterion profile $\rho_0(z)$ reduces
to a surface density coating the charged plane with a
density of ${n_0 / Z}$.  Therefore, according to PB theory, all of
the counterions collapse onto the charged plane at zero temperature.
\begin{figure}
\resizebox{2.5in}{2.5in}{\rotatebox{0}{\includegraphics{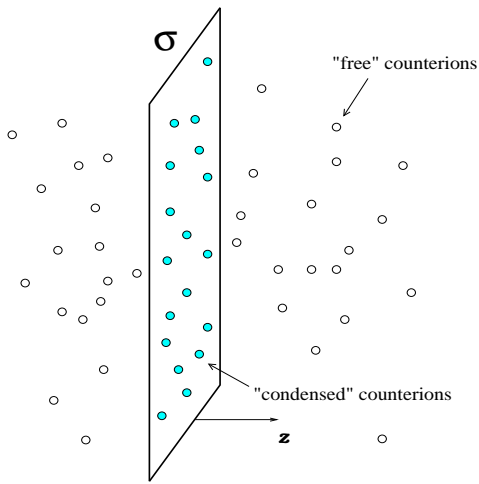}}}
\caption{The geometry of the problem.}
\label{geometry}
\end{figure}
However, for highly charged surfaces $Z^2 {l}_B \gg \lambda$,
the fluctuation corrections become so large that the solution
to the PB equation is no longer valid \cite{netz}.
To capture this regime in the spirit of the ``two-fluid'' model
 \cite{cond},
we divide the counterions into a ``free'' and a condensate
fraction.  The ``free'' counterions have the
usual PB 3D spatial distribution, while the ``condensed'' counterions
are confined to move only on the charged plane, as shown in Fig.
\ref{geometry}.  The free energy per unit area
for the condensed counterions with an average surface density $n_c$
can be written as \cite{2Dfree}\begin{eqnarray}
\beta f_{2d}(n_c) &=& n_c\left
\{ \ln[ n_c\,a^2] - 1 \right \}
\nonumber \\
&+& { 1 \over 2}\int {d^{2} {\bf q} \over (2 \pi)^{2}} \left \{
\ln \left [ 1 + \frac{1}{q\lambda_{D}} \right ] -
\frac{1}{q\lambda_{D}}
\right \},
\label{2Dplane}
\end{eqnarray}
where $\beta^{-1} = k_B T$, $a$ is the molecular size of the
counterions,
and $\lambda_{D} =  1/(2 \pi Z^2 n_{c})$ is
the 2D screening length.  The first term in Eq. (\ref{2Dplane}) is the
entropy and the second term arises from the 2D fluctuations.  Note that the
latter term is logarithmically divergent, which may be regularized by a microscopic
cut-off $\sim a$, yielding $\beta \Delta f_{2d}(n_c) \simeq  -{1 \over 8 \pi
\lambda_{D}^2}\ln(2\pi \lambda_{D}/a)$.  In addition, the condensate partially
neutralizes the charged plane, effectively reducing its surface charge density from $e n_0$
to $e n_R = e n_0 - Ze n_c$.  Thus, motivated by PB theory,
the free counterions can be modelled as an ideal gas confined to a slab
of thickness $\lambda_R \equiv  1 /(\pi l_B Z n_R)$ with a 3D
concentration of $c = n_R / (Z\lambda_R)$.  The fluctuation free
energy in this case may be estimated using the 3D Debye-H\"{u}ckel
theory: $\beta \Delta f = -\,\kappa_s^3/(12\pi)$ \cite{landau}, where $
\kappa_s^2 = 4 \pi Z^2 l_B c$ is the inverse square of the 3D screening length.
The free energy per unit area of the free counterions is then approximately given by\bb
\beta f_{3d}(n_c) \approx c\, \lambda_R  \left \{ \ln[ c\,a^3] - 1
\right \}  -
{\kappa_s^3 \over 12 \pi}\,\lambda_R.
\label{app3dfree}
\en
All the qualitative results, including the nature of the
condensation transition, follow straightforwardly from the analysis of
the total free energy: $f(n_c) = f_{2d}(n_c) + f_{3d}(n_c)$; minimizing $f(n_c)$
to find the fraction of condensed counterions, $n_c$, we obtain \bb
\ln \left [ \tau \over ( 1 - \tau )^2 \theta g \right ]+
{4 \over 3}\,g (1 - \tau) - \tau g \ln \left ( {\pi \over \tau \theta
g} \right )= 1,
\en
where the three dimensionless parameters: the order parameter $\tau \equiv Z n_c /n_0$,
the coupling constant $g \equiv Z^2 {l}_B / \lambda$, (where
$\lambda$ is the {\em bare} Gouy-Chapmann length), and the reduced
temperature $\theta \equiv {a \over Z^2 l_B}$, completely determine
the equilibrium state of the system.  It is easy to derive the
asymptotic solutions of the last equation corresponding to the
free, $\tau_1 \ll 1$, and condensed, $\tau_2 \approx 1$, state of
the counterions: $\tau_1  \simeq g\,\theta \exp \left[1-\frac{4}{3}\,g\right] $,
and $\tau_2 \simeq 1 - \left[ \pi \exp (1)\right] ^{-1/2}\left(
\frac{g \theta }{\pi } \right) ^{g - 1 \over 2}$, respectively.
For weak couplings $g \ll 1$, $\tau_1$ is the only consistent solution.
Thus, there is almost no condensed counterions
$\tau \ll 1$.  This is not surprising since PB theory is a
weak-coupling theory which becomes exact in the limit $T \rightarrow \infty$.
However, for high surface charge $g \gg 1$, where correlation effects becomes
important, the behavior of $\tau$ depends crucially on $\theta$.
In particular, for $\theta < \theta_c \approx 0.038$,
$\tau_1$ and $\tau_2$ are both consistent solutions corresponding
to the two minima of $f$, and thus a first-order transition takes place
when $f(\tau_1) = f(\tau_2)$, in which a large fraction of conterions
is condensed. This occurs at a particular value of the bare surface charge density
such that $g = g_0(\theta)$.  For an estimate, we take
$\theta = 0.02$ (divalent counterions at room temperature)
and obtain $g_0 \sim 1.7$, corresponding to $\sigma_c \sim e/10$
nm$^{-2}$.  However, for $\theta > \theta_c$ the behavior
of $\tau$ is completely different; in this regime, there is {\em no
phase transition} and the condensation occurs continuously.  Thus, the condensation
transition is similar to the liquid-gas transition, which has a line of
first-order transitions terminating at the critical point where a {\em
second-order} transition occurs.  If one takes $l_B \sim 10$ \AA, {\em e.g.}
room temperature, and $ a \sim 1\,$\AA, it follows from the definition
of $\theta$ that there is a critical value of counterion valence $Z_c =
\sqrt{a/(l_B \theta_c)} \simeq 1.62$, below which no first-order condensation transition is
possible. Therefore, divalent counterions behave {\em qualitatively} differently
from monovalent counterions at room temperature.

Clearly, this condensation picture may also be crucial to understanding
the attraction between two similarly charged plates, separated by
a distance $d$.  Recall that the total pressure of this system is
comprised of the mean-field repulsion and the correlated fluctuation
attraction \cite{fluct}. The repulsion comes
solely from the ideal gas entropy and it is proportional to
the concentration at the mid-plane:
$\Pi_0(d) = k_B T \rho_0(0) = 8 k_B T /( \ell_B \lambda_R^2)$ for
$ d < \lambda_R$ \cite{mean}.  The fluctuation-induced attraction
is $\Pi(d) =  -\,\alpha_0\,k_B T/d^{3}$ for $d > \lambda_D$, where
$\alpha_0 \approx 0.048$ \cite{fluct}.  Clearly, when
a large fraction of the counterions is ``condensed'',
the mean-field repulsion is greatly reduced.
Therefore, the attraction arising from correlated
fluctuations of the ``condensed'' counterions can overcome the
mean-field repulsion even for large distances.
Using the estimates in the last paragraph above, we find that
for {\em divalent} counterions and surface charge density of about one
unit charge per $\Sigma \sim 7\,\mbox{nm}^2$, the total pressure
becomes attractive at about $d \sim 10\,\,\mbox{nm}$; hence a
long-ranged attraction.  Of course, this estimate should be supplemented
by a more precise calculation for the system of two charged plates,
which is done below.

\section{Counterion Free Energy in The ``Two-Fluid'' Model}
\label{sec:twofluid}

Consider an overall neutral system consisting of counterions and
two charged surfaces separated by a distance $d$ immersed in an aqueous
solution.  The surface charged density on each plate is $\sigma_0 = e
n_0$.  We model the aqueous solution with a uniform dielectric constant
$\epsilon$.  This simplification allows us to study fluctuation and
correlation effects analytically.  In the spirit of the ``two-fluid''
model proposed in Ref. \cite{cond}, we divide the counterions into
a ``condensed'' and a ``free'' fraction.  The condensed counterions are
allowed to move only on the charged surfaces, while the free counterions
distribute in the space inside and outside the gap.
The number of the condensed counterion $n_c$ on each plate is
determined by minimizing the total free energy including fluctuation contributions.
Thus, our first task is to derive an expression for the total free energy of the system.
The electrostatic free energy for the whole system may be written as
\begin{widetext}
\begin{eqnarray}
\beta F_{el} &=& \sum_i\int d^2{\bf r}\, n_c^i({\bf r})
\left \{ \ln \left [ n_c^i({\bf r}) a^2 \right ] -  1 \right \}
+  \int d^3{\bf x}\, \rho({\bf x})
\left \{ \ln \left [ \rho({\bf x}) a^3 \right ] -  1 \right \} \nonumber \\
&+&  {Z^2l_B\over 2}\, \sum_{ij}
\int d^3{\bf x} \int d^3{\bf x}'\,
{ n_c^i({\bf r})\delta(z-z_i)\,n_c^j({\bf r}')\delta(z-z_j) \over | {\bf
x} - {\bf x}' | }  + {Z^2 l_B \over 2}\,
\int d^3{\bf x} \int d^3{\bf x}'\,{ \rho({\bf x}) \rho({\bf x}') \over | {\bf
x} - {\bf x}' | } \nonumber \\
&+& {Z l_B } \sum_{i} \int d^3{\bf x} \int d^3{\bf x}'\,
{ n_c^i({\bf r})\delta(z-z_i) [Z \rho({\bf x}') - n_f({\bf x}')] \over | {\bf x} - {\bf x}' | }
-  {Z l_B }\int d^3{\bf x} \int d^3{\bf x}'\,
{  \rho({\bf x})\,n_f({\bf x}') \over | {\bf x} - {\bf x}' |} \nonumber \\
&+& {l_B \over 2}\int d^3{\bf x} \int d^3{\bf x}'\,
{  n_f({\bf x}) n_f({\bf x}') \over | {\bf x} - {\bf x}' |},
\label{2dfree1}
\end{eqnarray}
\end{widetext}
where $a$ is the molecular size of the counterions,
$l_B = e^2 / (\epsilon k_B T)$ is the Bjerrum length, $Z$ is the
valence of the counterions,
and $z_1 = -d/2$ and $z_2 = d/2$ are the locations of the charged
surfaces. The first two terms in Eq. (\ref{2dfree1}) are the two-dimensional
entropy for the condensate and three-dimensional entropy for
the ``free'' counterions, respectively, and
the rest represent the electrostatic interactions of counterions in the
system. In Eq. (\ref{2dfree1}), the condensed counterions two-dimensional
density on the $i$-th plate is denoted by $n_c^i({\bf r})$,
the ``free'' counterions with 3D density by $\rho({\bf x})$, and
the external fixed charges arising from the surfaces by $n_f({\bf x}) =
n_0 \delta(z-d/2) +  n_0 \delta(z+d/2)$.  Within the Gaussian
fluctuation approximation, we assume that the 2D density of condensed counterions
has a spatially dependent fluctuation about a uniform mean:
$n_c^i({\bf r}) = n_c + \delta n_c^i({\bf r})$, and expand
Eq. (\ref{2dfree1}) to second order in $\delta n_c^i({\bf r})$.
Summing over all the 2D fluctuations of the condensed counterions, {\em i.e.}\[
e^{ - \beta {\cal H}_e}  =
\int {\cal D}\delta n_c^A({\bf r}) {\cal D}\delta n_c^B({\bf r})
e^{-\beta F_{el}},
\]
we obtain two terms in the effective free energy:  ${\cal H}_e = F_{2d}
+ {\cal H}_{3d}$.
The first term $F_{2d}$ is the free energy associated with the
condensed counterions which can be written as\begin{eqnarray}
\beta F_{2d} &=& 2 n_c\left \{ \ln[ n_c\,a^2] - 1 \right \} {\cal A}
+{1 \over 2} \ln \det \hat{{\bf K}}_{2d} \nonumber \\
&-& {1 \over 2} \ln \det[ -\nabla_{\bf x}^2],
\label{2D}
\end{eqnarray}
where $\hat{{\bf K}}_{2d}({\bf x}, {\bf y}) \equiv
\left [ - \nabla_{{\bf x}}^2 + {2 \over
\lambda_D}\,\sum_{\pm} \delta(z \pm d/2) \,\right ]\delta( {\bf x} -
{\bf y})$ is the 2D Debye-H\"{u}ckel operator and $\lambda_{D} =  1/(2\pi
Z^2 l_B n_{c})$ is the Debye screening length in 2-D.  The first term in Eq. (\ref{2D})
is the entropy and the second term arises from the 2D charge-fluctuations.  Note that
although this fluctuation term can be evaluated analytically \cite{fluct},
we write it in this abstract form for later convenience.

The second term ${\cal H}_{3d}$ is the electrostatic free energy
for the ``free'' counterions, taking into account of the presence
of the fluctuating condensate; to within an additive constant, it may
be written as
\begin{eqnarray}
\beta {\cal H}_{3d} &= &\int d^3{\bf x}\,
\rho({\bf x}) \left \{ \ln \left [ \rho({\bf x}) a^3 \right ] -  1
\right \} \nonumber \\
&+& {1 \over 2}\,\int d^3{\bf x} \int d^3{\bf
x}'\,\rho({\bf x})
G_{2d}({\bf x}, {\bf x}')\rho({\bf x}') \nonumber \\
&-&  \int d^3{\bf x}\,\phi({\bf x})\,\rho({\bf x}),
\label{energy}
\end{eqnarray}
where $\phi({\bf x}) \equiv \int d^3{\bf x}'\,Z^{-1}\,
G_{2d}({\bf x}, {\bf x}')\,n_R({\bf x}')$ is the ``renormalized''
external field
arising from the charged plate.  From Eq. (\ref{energy}), we can see
that
the presence of the condensate modifies the electrostatics of
the free counterions in two ways.  First, the condensate partially
neutralizes
the charged surfaces, effectively reducing the surface charge density
from $e n_0$
to $e n_R = e(n_0 - Z n_c)$.  Second, their fluctuations
renormalize the electrostatic interaction of the system;
thus, instead of the usual Coulomb potential, the free counterions and
the charged surfaces interact via the interaction
$G_{2d}({\bf x},{\bf x}')$, which is the inverse (the Green's function)
of the
2D Debye-H\"{u}ckel operator $\hat{{\bf K}}_{2d}$:\bb
\left [ - \nabla_{{\bf x}}^2 + {2 \over \lambda_D}\,\sum_{\pm} \delta(z
\pm d/2) \right ]
G_{2d}({\bf x}, {\bf x}') = \ell_B \delta({\bf x} - {\bf x}'),
\label{2dDebye}
\en
where we have defined, for convenience, a reduced Bjerrum length by
$\ell_B \equiv 4\pi Z^2 l_B$.  In Eq. (\ref{2dDebye}), the second term
in the bracket takes into account of the fluctuating ``condensate''.
Hence, in the limit
$n_c \rightarrow 0$ or $\lambda_{D} \rightarrow \infty$,
$G_{2d}({\bf x},{\bf x}')$ reduces to the usual Coulomb interaction
$G_{0}({\bf x},{\bf x}') = \ell_B / |{\bf x}-{\bf x}'|$.

After a Hubbard-Stratonovich transformation \cite{hubstr},
the grand canonical partition function for the ``free'' counterions,
characterized by the interaction energy in Eq. (\ref{energy}),
can be mapped onto a functional integral representation:
${\cal Z}_{\mu}[\phi] = {\cal N}_0\,\int {\cal D}\psi\,
e^{- {\cal S}[\psi,\phi]}$ with an action \cite{samu}
\begin{eqnarray}
{\cal S}[\psi,\phi] &=& \int {d^3{\bf x} \over {\ell}_B }\,\left
\{\,
{1 \over 2}\,\psi({\bf x}) [\, - \nabla^2 \,] \psi({\bf x}) - \kappa^2
e^{i \psi({\bf x}) + \phi({\bf x})\,}
\vphantom{{1 \over \lambda_D}\,\sum_{\pm}\delta(z\pm d/2)\, [\psi({\bf
x})]^2} \right. \nonumber \\
&+& \left. {1 \over \lambda_D}\,\sum_{\pm}\delta(z\pm d/2)\, [\psi({\bf
x})]^2 \right \},
\label{action}
\end{eqnarray}
where $\psi({\bf x})$ is the fluctuating field, $\kappa^2 =
e^{\mu}{\ell}_B/a^3 $,
$\mu$ is the chemical potential and ${\cal N}_0^{-2} \equiv \det
\hat{{\bf K}}_{2d}$ is
the normalization factor.  The minimum of the action, given by
$\left. { \delta {\cal S} \over \delta \psi({\bf x}) }\right |_{\psi
=\psi_0} = 0$, defines the saddle-point equation for $\psi_0({\bf x})$\begin{eqnarray}
- \nabla^2 [i \psi_0( {\bf x})] &+& { 2 \over \lambda_D
}\,\sum_{\pm}\delta(z\pm d/2)
\,[i \psi_0( {\bf x})] \nonumber \\
&&\phantom{++++} +\kappa^2 e^{i \psi_{0}({\bf x}) + \phi({\bf x})} = 0.
\label{saddle}
\end{eqnarray}
This saddle-point equation is equivalent to the PB equation by defining
the mean-field potential $\varphi({\bf x}) = - i \psi_{0}({\bf x}) -
\phi({\bf x})$, which is solved below in Sec. \ref{sec:twoplate}.
To obtain the free energy for the
free counterions on the mean-field level, we note that it is related
to the Gibbs potential $\Gamma_0[\phi] \equiv  {\cal S}[\psi_0, \phi]$
by a Legendre transformation:\bb
\beta F_{3d}^0(n_R) =  \Gamma_0[\phi] + \mu \int d^3{\bf x}\,
\rho_0({\bf x}),
\label{mean}
\en
where $\rho_0({\bf x})$ is the mean-field free counterion density given
by\bb
\rho_0({\bf x}) = \left ({\kappa^2 / \ell_B}\right )\,
e^{i \psi_0({\bf x}) + \phi({\bf x})\,}.
\en

To capture correlation effects, we must also include fluctuations of
the ``free'' counterions, thereby treating the ``free'' and ``condensed''
counterions on the same level.  To this end, we expand the action
${\cal S}[\psi,\phi]$ about the saddle-point $\psi_0({\bf x})$ to second order
in $\Delta \psi({\bf x}) = \psi({\bf x}) -\psi_0({\bf x})$:
\begin{widetext}
\begin{eqnarray}
{\cal S}[\phi,\psi] = {\cal S}[\phi,\psi_0]
+ {1 \over 2}\,\int d^3{\bf x}\,\int d^3{\bf y}\,
\Delta\psi({\bf x})\,\hat{{\bf K}}_{3d}({\bf x}, {\bf
y})\,\Delta\psi({\bf y})+ \cdots,
\nonumber
\end{eqnarray}
where the differential operator\begin{eqnarray}
\hat{{\bf K}}_{3d}({\bf x}, {\bf y}) \equiv
\left [ - \nabla_{{\bf x}}^2\,+\,{2 \over
\lambda_D}\,\sum_{\pm}\delta(z \pm d/2)\,+ \,
\kappa^2 e^{i \psi_0({\bf x}) + \phi({\bf x})\,} \right ] \delta( {\bf x} - {\bf y}),
\label{K}
\end{eqnarray}
\end{widetext}
is the second variation of the action ${\cal S}[\psi,\phi]$.  Note
that the linear term in $\Delta \psi({\bf x})$ does not contribute to
the expansion since $\psi_0({\bf x})$ satisfies the saddle-point equation Eq.
(\ref{saddle}).  Performing the Gaussian integrals in the functional integral,
we obtain an expression for the change in the free energy due to fluctuations of the
free counterions:\bb
\beta \Delta F_{3d}= {1 \over 2} \ln \det \hat{{\bf K}}_{3d}  -
{1 \over 2} \ln \det \hat{{\bf K}}_{2d},
\label{3D}
\en
where the second term comes from the normalization factor ${\cal N}_0$.
Note that the second term in Eq. (\ref{3D}) partially cancels the
fluctuation contributions to the free energy of the condensed counterions in Eq.
(\ref{2D}). Thus, combining Eqs. (\ref{2D}), (\ref{mean}), and (\ref{3D}) together,
the total free energy of the system can be expressed as\begin{eqnarray}
\beta F(n_c) &=&   2 n_c\left \{ \ln[ n_c\,a^2] - 1 \right \} {\cal A}
+ \beta F_{3d}^0(n_R) \nonumber \\
&+& {1 \over 2} \ln \det \hat{{\bf K}}_{3d}
- {1 \over 2} \ln \det[ -\nabla_{\bf x}^2].
\label{totalfree}
\end{eqnarray}
This is the main result of this paper, from which all the equilibrium
quantities can be calculated.  It says that the free energy
of the counterions is simply a sum of the mean-field free energy and a
fluctuation energy term.  Note that the latter term contains couplings
among the fluctuations of the free and the condensed counterions.
Finally, we stress that the derivation presented here is rather general
and may apply to other physical systems as well.

\section{Interaction between Two Similarly Charged surfaces}
\label{sec:twoplate}

In this section, we employ the framework of counterion condensation
derived in Sec. \ref{sec:twofluid} to study the interaction of
two charged surfaces with free counterions and condensed counterions
fluctuating on each of them.  In Sec. \ref{subsec:meanfield},
an expression is derived for the total pressure, which takes into
account the total fluctuations of the counterions.  In Sec.
\ref{subsec:freeenergy}, we discuss the behavior of the total pressure
and the equilibrium state of the system as characterized by the fraction
of condensed counterion $\tau$ which is determined by the minimum of
the total free energy Eq. (\ref{totalfree}).

\subsection{Mean-Field Theory and Fluctuation Corrections}
\label{subsec:meanfield}

The free counterion density on the mean-field level can be obtained by
solving Eq. (\ref{saddle}).  Defining the mean-field potential by
$\varphi({\bf x}) = - i \psi_{0}({\bf x}) -
\phi({\bf x})$, the saddle-point equation becomes\begin{eqnarray}
{d^2\varphi(z) \over dz^2} + \kappa^2 e^{-\varphi(z)} &=&
{n_R \ell_B \over Z}\,\sum_{\pm} \delta(z \pm{ d /2}) \nonumber \\
&+& { 2 \over \lambda_D } \sum_{\pm} \delta(z \pm{ d /2}) \varphi(z),
\label{pb}
\end{eqnarray}
where $e n_R = e(n_0 - Z n_c)$ is the {\em renormalized} surface charge
density of the plates.  Note that Eq. (\ref{pb}) looks similar to
the mean-field PB equation.  Indeed, the solution to Eq. (\ref{pb})
is exactly the same as the PB solution provided that we impose the
boundary condition: $\varphi(\pm\,d/2) = 0$.  The solution
reads\begin{eqnarray}
\varphi_<(z) &=&   \ln { \kappa^2  \cos^2 ( \alpha z) \over
2\,\alpha^2 },\,
\mbox{$|z| \leq d/2$}, \\
\varphi_>(z) &=&   2 \ln \left [ 1 + {\kappa  \over \sqrt{2}}
\left ( |z| - {d /2} \right )\right ],\,
\mbox{$|z| \geq d/2$},
\end{eqnarray}
and the counterion density is given by\begin{eqnarray}
\rho_0 e^{-\varphi_<(z)} &=& (2 \alpha^2 / \ell_B) \sec^2(\alpha z),\,\mbox{$|z| \leq d/2$},\\
\rho_0 e^{-\varphi_>(z)} &=&
{2 / \ell_B\, \over \left (\, |z| - {d /2} + \xi \, \right )^{2}},\,
\mbox{$|z| \geq d/2$},
\end{eqnarray}
where $\xi \equiv \sqrt{2}/ \kappa$ and $\alpha$ is determined
from the boundary conditions on the electric field:
$\left. { \partial_z \varphi_> } \right|_{d/2} -
\left. { \partial_z \varphi_<  } \right|_{d/2}
=  {n_R \ell_B / Z}$ and $\varphi(\pm\,d/2) = 0$;
they lead to a transcendental equation for $\alpha$:\bb
\,\alpha \lambda_R \tan \left ( \alpha d/2 \right ) =  1 - ( \alpha
\lambda_R /2)^2,
\label{tran}
\en
where $\lambda_R = 4 Z /(\ell_B n_R)$ is the {\em renormalized}
Gouy-Chapmann length.
Physically, $\alpha^2$ is proportional to the free counterion density
at the mid-plane $\rho_0(0)$. In addition, $\kappa$ is related to the zeros of
the potential, {\em i.e.} $\varphi(\pm\,d/2) = 0$:\bb
\kappa^2 = 2 \alpha^2 \sec^2 \left (\alpha d/2 \right)
= {2\,( 1 + b^2 )^2/ \lambda_R^2},
\label{trans}
\en
where we have defined $b \equiv \alpha \lambda_R /2$.  The
asymptotic behaviors for $b$ as determined by the relation Eq.
(\ref{tran}) are $b \sim 1/d$ as $d \rightarrow \infty$ and
$b \sim 1$ as $d \rightarrow 0$.  We note that the latter behavior is
distinct from the case of two impenetrable charged hard walls \cite{sam}.

The mean-field free energy per unit area
for the free counterions, i.e., the first two terms in Eq.
(\ref{totalfree}), can be easily calculated by using Eq. (\ref{mean}),\begin{eqnarray}
\beta f_0 &=&  2 n_c\left \{ \ln[ n_c\,a^2] - 1 \right \} + {2 n_R
\over Z}\left \{
\ln \left ( { n_R a^3 \over 2 Z \lambda_R}\right ) - 1 \right \}
\nonumber \\
&+& { 4 n_R \over Z} \ln \left [ 1 + \left (\alpha \lambda_R /  2
\right )^2 \right ]
+ { 2 \alpha^2 d \over \ell_B},
\label{fmf}
\end{eqnarray}
where we have made used of the fact that the chemical potential is
given by $\mu = \ln \left ( \kappa^2 a^3/\ell_B \right)$.
The first two terms in Eq. (\ref{fmf}) represent the entropy per unit
area of condensed and free counterions,
respectively.  The last two terms describe the interacting free energy
for the surfaces.  Using the general formula for the pressure:
$\Pi_0(d) = -\,{\partial f_{0}(d) \over \partial d}$,
we obtain the mean-field pressure between the surfaces: $\Pi_0(d) =
+\,{ 2 \alpha^2 / \ell_B}$.  We note that at the mean-field level,
the pressure comes solely from the ideal gas entropy of the ``free'' counterions,
and it is proportional to their density at the midplane, a standard result.
However, in contrast to the standard PB theory, the pressure now depends on
the order parameter $\tau \equiv Z n_c /n_0 $.  Thus, if there were a large fraction of
condensed counterions, $\tau \simeq 1$, the mean-field repulsion would be
drastically reduced.

Next, we compute the pressure arising from the counterion fluctuations.
Recall that the expression for the change in the free energy arising
from fluctuations of the counterions is given by the last two terms in Eq.
(\ref{totalfree})\bb
\beta \Delta {\cal F} =  { 1\over 2}\ln \det \hat{{\bf K}}_{3d}
- {1 \over 2} \ln \det \left [ - \nabla_{\bf x}^2  \right ],
\label{fluctf}
\en
where the operators $\hat{{\bf K}}_{3d}$ is defined in Eq. (\ref{K})
with\bb
\kappa^2 e^{-\varphi(z)\,} =
2 \alpha^2 \sec^2(\alpha z)\,\Theta(z) +
{ 2 \,\tilde{\Theta}(z)\over \left (\, |z| - d/2 + \xi \,\right )^{2}},
\label{density}
\en
where $\xi \equiv \lambda_R/(1+b^2)$, $\Theta(z) = \theta(z+d/2) -
\theta(z-d/2) = 1,$ if $|z| \leq d/2$ and zero, otherwise,
and $\tilde{\Theta}(z)  = 1- \Theta(z)$.  The derivative of $\Delta
{\cal F}$ with respect to distance $d$ can be straightforwardly calculated by
making use of the exact identity: $\delta \ln \det \hat{{\bf X}} =
{\mbox{Tr}}\,\hat{\bf X}^{-1}\,\delta\,\hat{\bf X}$,\begin{eqnarray}
&& {\partial \beta \Delta {\cal F} \over \partial d } =
{ 1 \over 2 \ell_B} \int d^{3}{\bf x}\, G_{3d}({\bf x},{\bf
x}) \times \nonumber\\
&& \hphantom{+++}
{\partial \over \partial d } \left [ {2 \over \lambda_D }\,\sum_{\pm}\,\delta(z \pm d/2)
+\, \kappa^2 e^{-\varphi(z)} \right ],
\label{expressure}
\end{eqnarray}
where $G_{3d}({\bf x},{\bf x}')$ is the Green's function of the
operators $\hat{{\bf K}}_{3d}$ satisfying \begin{eqnarray}
&& \left [\,- \nabla_{\bf x}^2 + {2\over \lambda_D} \sum_{\pm} \delta(z
\pm d/2) + \, \kappa^2 e^{-\varphi(z)}\, \right ]G_{3d}({\bf x},{\bf x}') \nonumber \\
&& \hphantom{++++++++++++++} =\ell_B\,\delta({\bf x} - {\bf x}').
\label{greenwalls}
\end{eqnarray}
An explicit derivation of $G_{3d}({\bf x},{\bf x}')$ and the pressure
arising from fluctuations Eq. (\ref{expressure}) are detailed in
Appendix \ref{app:derivation}.
The final result for the pressure can be written as\begin{eqnarray}
\beta\Pi(d) &=& - { 1\over {\cal A}} { \partial \Delta {\cal F} \over
\partial d }
= -\,\int {d^{2}{\bf q} \over (2 \pi)^2}\,
{ q {\cal M}^2(q) \over 1- {\cal M}^2(q) } \nonumber \\
&-& \,{\alpha^2 \over \lambda_R }\,
{ (1 + b^2)\,(\,{\cal I}_2 - {\cal I}_3\,)
\over 2 + {\,d\,\over \lambda_R}\,(1 + b^2\,)\,},
\label{pressurefin}
\end{eqnarray}
where ${\cal M}(q)$ is defined in Eq. (\ref{M}), ${\cal I}_2$ and
${\cal I}_3$ are two dimensionless integrals defined in the Appendix
\ref{app:derivation} by Eqs. (\ref{I2}) and (\ref{I3}), respectively.

We note that the fluctuation pressure is purely attractive; thus,
fluctuations lower the free energy.  Although the expression Eq. (\ref{pressurefin})
looks complicated, each term, however, has a simple physical interpretation.
The first term in Eq. (\ref{pressurefin}) is the
pressure arising from counterion fluctuations near the surfaces.  In
fact, if all of the counterions is condensed $\tau =1$, we observe that
${\cal M}(q)$ in Eq. (\ref{M}) becomes ${\cal M}(q) = -  e^{-qd} /( 1 + q \lambda_D)$
and that the only contribution to the pressure in Eq. (\ref{pressurefin})
is the first term, which becomes in this limit\bb
\beta\Pi(d) = - \int {d^{2}{\bf q} \over (2 \pi)^2}\,{q \over e^{2qd}
(1+q \lambda_D )^2 - 1}.
\en
This expression is exactly the pressure derived in Ref. \cite{fluct},
arising from 2D fluctuations of the counterions. It scales like $\Pi(d)
\sim - 1/d^3$ for large distances.

The second term in Eq. (\ref{pressurefin}) may be interpreted as
the coupling between the counterions near the surfaces and those in the
bulk. This can be seen by considering the asymptotic behavior of the pressure
for large $d$ in the no condensate limit $\tau =0$, {\em i.e.}
the fluctuation corrections to the PB pressure.  In this case, in
addition to the usual $d^{-3}$ scaling law arising from counterion fluctuations
near the surfaces, the second term in Eq. (\ref{pressurefin}) contributes a term,
which scales as $ \sim d^{-3}\,\ln \left ( d/\lambda \right )$
in the large $d$ limit.  Therefore, the pressure\bb
\Pi(d)  \sim - { 1 \over d^3 } -  { 1 \over d^3 } \ln \left ( d /
\lambda \right ),
\en
contains a logarithmic term, which dominates the $d^{-3}$ term for
large distances. This term has been obtained by several authors previously
 \cite{ass,fluct2} and in particular, Ref. \cite{fluct2} shows that this term arises
physically from the coupling between counterions near the surfaces
and those in the bulk.  Therefore, Eq. (\ref{pressurefin}) recovers the
PB limit $\tau = 0$ and the 2D limit $\tau = 1$ as special cases.  Although the
fluctuation corrections to the PB ($\tau = 0$) pressure have been considered
previously, we stress that Eq. (\ref{pressurefin}) is a generalization
which allows for counterion condensation and may apply to other physical
situations, such as ions absorption.

Combining with the mean-field pressure, we obtain the total
pressure\begin{eqnarray}
\beta \Pi_{tot}(d) &=& { 2 \alpha^2 \over \ell_B} \left \{ 1 -
{\ell_B \over 8 \pi \lambda_R } { (1 + b^2)\,
(\,{\cal I}_2 - {\cal I}_3\,) \over 2 + {\,d\,\over \lambda_R}\,(
1 + b^2\,)\,}\right \} \nonumber \\
&-& \int {d^{2}{\bf q} \over (2 \pi)^2}\,{ q {\cal M}^2(q) \over 1- {\cal
M}^2(q) }.
\label{totalpressure}
\end{eqnarray}
The behavior of the total pressure depends on the coupling constant
$g \equiv Z^2 {l}_B / \lambda$ and the fraction of condensed
counterions
$\tau \equiv Z n_c /n_0$.  For $g \ll 1$ and $\tau \ll 1$, the
fluctuation corrections
are small and the total pressure $\Pi_{tot}(d)$ is controlled by
the mean-field repulsion.  On the other hand, for $\tau \sim 1$
the mean-field repulsion is greatly reduced and the fluctuation
attraction can overcome
the repulsion at finite distances.  Furthermore, for $g \sim 1$ the
short distance behavior is highly sensitive to $\tau$:
Even a very small number of condensed counterions would turn the total
pressure, otherwise repulsive for $\tau =0$, into attractive for short
distances.
For $g \gg 1$, the fluctuation attraction becomes dominant at short
distances
even when there is no condensate, and the effect of finite $\tau$ is to
push the attractive
region out to a larger length scale.  Hence, if there is
sufficient number of condensed counterions, the pressure is
attractive even for large distances.  Our next task is to
determine the fraction of condensed counterions $\tau$ as
determined by the minimum of the total free energy.

\subsection{Equilibrium Properties}
\label{subsec:freeenergy}

The equilibrium state of the system is determined by minimizing the
total free energy
with respect to the order parameter $\tau$.  Therefore, we
need to evaluate the derivative of the total free energy Eq.
(\ref{totalfree}) with respect to $\tau$.  Let us first consider the
mean-field contribution.
Explicitly differentiating Eq. (\ref{fmf}), we obtain\begin{eqnarray}
{\partial \beta f_0 \over \partial \tau} &=& {2 n_0 \over Z}\,\ln (
2 \lambda / a ) - {2 n_0 \over Z} + {2 n_0 \over Z}\, \ln {\tau
\over (1- \tau)^2 } \nonumber \\
&-& {2 n_0 \over Z} \ln \left [ 1 + b^2 \right ],
\label{meanfieldd}
\end{eqnarray}
where $\lambda = 4Z/(\ell_B n_0)$ is the ``bare'' Gouy-Chapman
length.  To obtain the fluctuation contributions, we again
make use of the exact identity: $\delta \ln \det \hat{{\bf X}} =
{\mbox{Tr}}\,\hat{\bf X}^{-1}\,
\delta\,\hat{\bf X}$ to evaluate the derivative
of the fluctuation free energy in Eq. (\ref{fluctf}),\begin{eqnarray}
{\partial \beta \Delta {\cal F} \over \partial \tau} &=&
{ 1 \over 2 \ell_B} \int d^{3}{\bf x}\, G_{3d}({\bf x},{\bf x})\,\times
\nonumber\\
&&{\partial \over \partial \tau } \left [{2\over \lambda_D}
\,\sum_{\pm}\,\delta(z \pm d/2) + \kappa^2 e^{-\varphi(z)}
\right ].
\label{deriF}
\end{eqnarray}
This expression can be explicitly evaluated using similar techniques
outlined in Appendix \ref{app:derivation} and the result is given
by\begin{eqnarray}
{ 1 \over {\cal A}}\,{\partial \beta \Delta {\cal F}
\over \partial \tau}
&=& {4 (1-\tau) \over \lambda^2 } \left [{1 \over 2}\,{\cal I}_1 + { 1
+ b^2 \over 2}\,{\cal I}_3
\vphantom{{4 (1-\tau) \over \lambda^2 }{2 b^2\,(\,{\cal I}_2 - {\cal
I}_3\,)
\over 2 + {\,d\,\over \lambda_R\,}(1 + b^2\,)}} \right. \nonumber \\
& & \hphantom{{1 \over 2}\,{\cal I}_1 +++} + \left. {2 b^2\,(\,{\cal
I}_2 - {\cal I}_3\,)
\over 2 + {\,d\,\over \lambda_R\,}(1 + b^2\,)} \right ],
\label{fluctd}
\end{eqnarray}
where ${\cal I}_2$ and  ${\cal I}_3$ are given in Eqs. ({\ref{I2}) and
(\ref{I3}),
respectively, and ${\cal I}_1$ is defined by\bb
{\cal I}_1[d/ \lambda_R] \equiv  \int{d^2{\bf q} \over
(2 \pi)^2 }\,{4 \pi \lambda_R \over 2 q }\, \left [\, {\cal G}(d/2) -
1 + {\cal L}(q)\, \right ],
\label{I1}
\en
where ${\cal L}(q)$ and ${\cal G}(d/2)$ are defined by Eqs. (\ref{L})
and (\ref{G}), respectively, in Appendix \ref{app:derivation}.
Note that ${\cal I}_1[d/ \lambda_R]$ is logarithmically divergent
(see Appendix \ref{app:derivation}, Eq. (\ref{I1fin})), as
in the case for 2D Debye-Huckel theory, which may be regularized by a
microscopic cut-off, chosen to be the size of the counterion $a$.
Finally, using Eqs. (\ref{meanfieldd}) and (\ref{fluctd}), the root of
the free energy $\partial F(\tau)/ \partial \tau =0$ can be
determined numerically.  For example, the case of an isolated charged
plate can be obtained by taking the limit $d \rightarrow \infty$ in Eqs.
(\ref{meanfieldd}) and (\ref{fluctd}),  which leads to the following transcendental
equation\begin{eqnarray}
1 &+& \ln \left  ( g \theta /2\, \right ) + \ln { (1- \tau)^2 \over \tau }
\nonumber \\
&+& { g  }\,\int_0^{x_c}\,x\,dx {  1 + 2 \gamma ( 1 + x) \over ( 1 +
x)[ 1 + (\gamma + x)(1 + x)] } =0,
\end{eqnarray}
where $x_c = {2\pi \over
(1-\tau)g \theta}$ is the microscopic cut-off.  The analysis of
this equation gives all the features mentioned in Sec.
\ref{sec:condensation}.  As a consistency check, it can be verified
that in the limit $d \rightarrow 0$, Eq. (\ref{fluctd}) gives the
fluctuation free energy for an isolated charged surface but with
twice of the surface charge density $2 e n_0$.

\section{Results and Discussion}
\label{sec:results}

\begin{figure}
\resizebox{2.5in}{2in}{\rotatebox{0}{\includegraphics{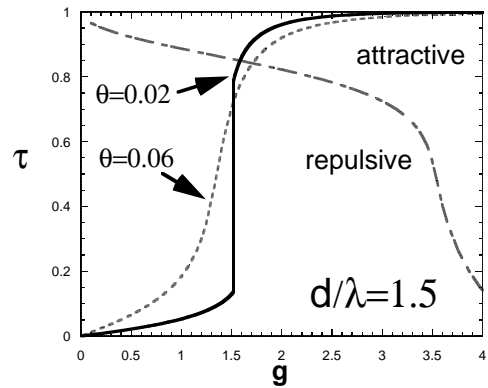}}}
\caption{The fraction of
condensed counterions $\tau \equiv Z n_c/n_0$ as
a function of $g \equiv Z^2 l_B/\lambda$ for different values of
$\theta = {a \over Z^2 l_B}$. At low surface charge $g \ll 1$,
the counterion distribution is well described by PB theory since
$\tau \ll 1$.  However, at high surface charge, correlation effects
leads to a large fraction of counterion condensed. The condensation
is first-order for $\theta < \theta_c$ and continuous for
$\theta > \theta_c$.  The critical point is at $\theta_c \sim 0.017$,
$g_c \sim 1.23$, and $\tau_c \sim 0.43$.}
\label{orderparp}
\end{figure}

Let us first discuss the behavior of the order parameter $\tau$ at
a fixed separation $d$ between the charged surfaces, as
summarized in Fig. \ref{orderparp}. The behavior of
$\tau$ as a function $g$ at a finite distance $d$ is qualitatively
identical to the case of infinite separation. For weak couplings $g \ll 1$,
there is small but finite number of condensed counterions but the
total pressure remains repulsive. For sufficiently high $g \sim
1$, the condensation proceeds continuously for $\theta > \theta_c$
and via a first-order phase transition for $\theta < \theta_c$ at a
particular value of the coupling constant $g_0(d,\theta)$. We note that
in this regime, the number of counterion condensation becomes
significant. This implies that the mean-field repulsion is
drastically reduced and the correlated attractions can overcome
the repulsion at finite distance. For $\theta \sim 0.02$, roughly
corresponding to divalent counterions at room temperature, we find that
the onset of the attraction occurs at $g \sim 1.6$ or surface charge of
about one charge per $10$ nm$^2$ at a distance $d = 1.5 \lambda \sim 40$ \AA.
These numbers are order of magnitude consistent with computer
simulations \cite{Guldbrand}.

\begin{figure}
\resizebox*{2.5in}{2in}{\rotatebox{0}{\includegraphics{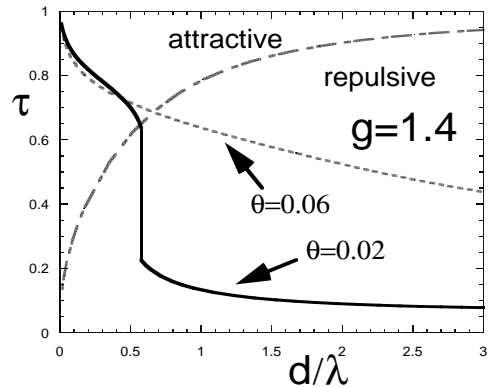}}}
\caption{The fraction of
condensed counterions as a function of distance.}
\label{phase1}
\end{figure}

Next, we discuss counterion condensation and the total pressure of
system as a function of distance.  Note that this scenario is more physically
relevant, since surface force experiments usually vary the distance between
charged surfaces rather than changing their surface charge densities.  For low surface
charges $g \ll 1$, as shown in Fig. \ref{smallg}, the counterion condensation is
continuous as a function of separation and the fraction of condensed counterion
remains small but finite.  We note that $\tau$ generally increases as the
distance of two surfaces decreases, but remains fairly constant up to $\lambda$.
This is not surprising since there is an entropy loss of the free counterions due to
confinement.  However, the pressure remains repulsive and shows little
difference from the PB pressure profile, as expected.

\begin{figure}
\resizebox*{2.5in}{2in}{\rotatebox{0}{\includegraphics{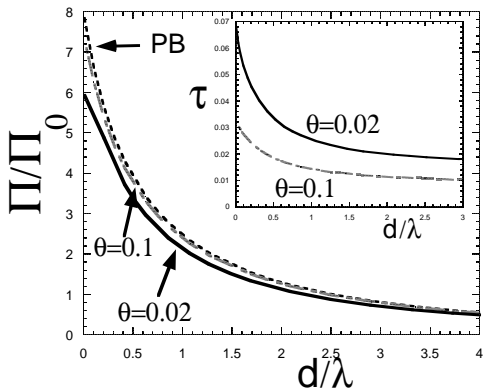}}}
\caption{The pressure profile for monovalent ($\theta =0.1$)
and divalent ($\theta =0.02$) counterions in the case of low
surface charges $g$.}
\label{smallg}
\end{figure}

\begin{figure}
\resizebox*{2.5in}{2in}{\rotatebox{0}{\includegraphics{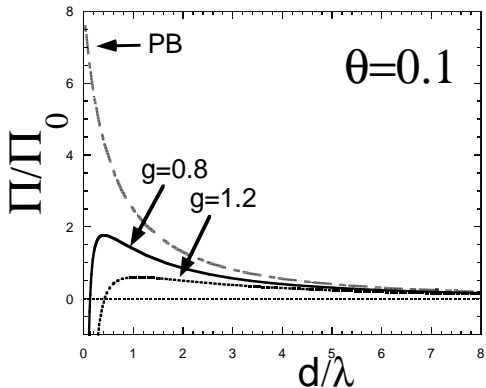}}}
\caption{The pressure profile for monovalent counterions
in the case of moderate coupling $g \sim 1$.}
\label{premono}
\end{figure}

For sufficiently high coupling $g \sim 1$, we have several
interesting regimes depending on the reduced temperature $\theta$ (see
Fig. \ref{phase1}).  For $\theta > \theta_c$, the counterions condense {\em continuously}
as the separation $d$ decreases and the pressure of the system remains
repulsive down to very short distances, as shown in Fig. \ref{premono}, where we
have plotted the pressure profile for monovalent counterions ($\theta = 0.1$)
for different values of $g$.  Note also that there is still a
repulsive barrier, which decreases with increasing $g$, while the range
of the attraction is shifted to larger separations.   For $g =1.2$, corresponding
to a surface charge density of about one charge per $\Sigma \sim 300$
\AA$^2$, the total pressure becomes attractive at about $d \sim 10$ \AA.
It should be noted that in real experimental settings, other strong
repulsive force, such as hard-core or hydration force, that we have not taken into
account in our model, may become important and may overwhelm this correlated attraction at
length scale less than $\sim 20\,\,\mbox{\AA}$ \cite{nature}.
This may explain why attraction is difficult to observe
experimentally for {\em monovalent} counterions. Moreover, the pressure
profile for large separations is similar to that of the PB theory,
except with a renormalized or effective surface charged density.
Indeed, it is known experimentally that in order to fit
experimental data to the PB theory, it is necessary to use an
effective surface charge, which is always lower than the actual
surface charge density \cite{nature}. Therefore, this counterion
condensation picture provides a possible scenario in which this
phenomenon can be accounted for theoretically, without invoking
charge regulation mechanism.

\begin{figure}
\resizebox*{2.5in}{2in}{\rotatebox{0}{\includegraphics{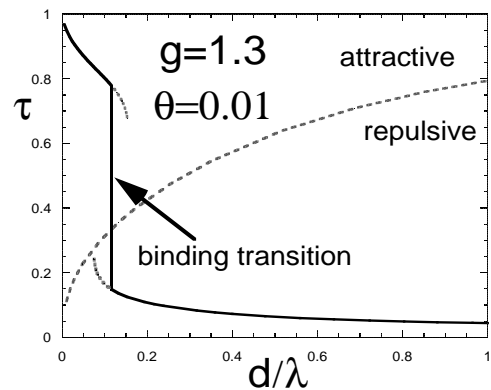}}}
\caption{First-order binding transition: for the case of trivalent
counterion ($\theta =0.01$), the number of condensed counterions
($\tau$) exhibits a discontinuous jump at a particular distance.}
\label{phase2}
\end{figure}

\begin{figure}
\resizebox*{2.5in}{2in}{\rotatebox{0}{\includegraphics{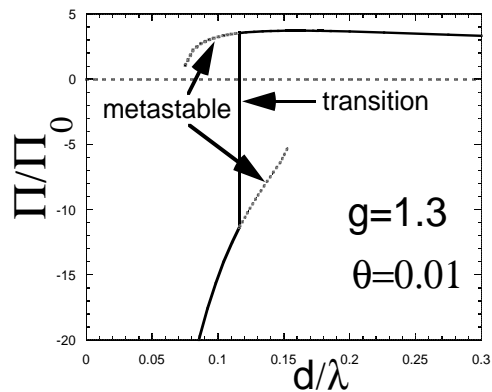}}}
\caption{The presure profile for the first-order binding transition.}
\label{preless}
\end{figure}

However, for $\theta  < \theta_c$ the behavior of the order parameter
$\tau$ and the total pressure of the system are qualitatively different (see
Fig. \ref{phase1}).  We find that there is a range in the coupling constant:
$g_{\infty}(\theta)/2 < g < g_{\infty}(\theta)$,
in which the order parameter displays a finite jump at a
particular separation $d_0(g,\theta)$, and the counterion
condensation is first-order as a function of the separation $d$.
Here, $g_{\infty}(\theta)$ denotes the coupling constant at which the
first-order counterion occurs at infinite separation, {\em i.e.} an isolated charged plate
(see Sec. \ref{sec:condensation} and Ref. \cite{cond}).  Thus, in the
limit $g \rightarrow g_{\infty}(\theta)$, we must have $d_0(g,\theta)
\rightarrow \infty$, since the system is composed of two isolated charged plates.  On the
other hand, we have $d_0(g,\theta) \rightarrow 0$ as $g \rightarrow
g_{\infty}(\theta)/2$, because this limit corresponds to a single
charged plate with twice of the surface charge density, {\em i.e.
$\sigma = 2e n_0$}.  This striking behavior of the order
parameter has interesting implications for the interaction for the
system.  Indeed, for sufficiently short distances, we find that
the first-order counterion condensation spontaneously can take the
system from the repulsive to the attractive regime, resulting in a novel
first-order binding transition.  This is illustrated in Fig. \ref{phase2} for the case of
trivalent counterions at room temperature $\theta = 0.01$ at $g =1.3$,
corresponding to a surface density of one charge per $\Sigma \sim 70$ nm$^2$.
(For trivalent counterions, the first-order counterion condensation occurs
in the range of $0.9 < g < 1.8$ \cite{cond}.)  The corresponding pressure
profile is plotted in Fig. \ref{preless}, which shows that the binding transition occurs at about
$d \sim 10$ \AA.  We note that an interesting consequence of this first-order
binding transition is the existence of metastable states, which may have important
manifestations in surface force experiments. It is easy to imagine that
the system can be trapped in different metastable states, and
therefore, hystersis may occur as the two surfaces are pushing in and pulling out
again. Indeed, there are some experimental support for this behavior for
multivalent counterions in similar systems \cite{avi}.  It is important to
emphasize that this interesting behavior is not included in the mean-field PB theory.
Note also that this first-order binding transition can only take place at short
distances.  This is because $d_0(g,\theta)$
generally increases with increasing $g$, and eventually when $g$ is
near $g_{\infty}(\theta)$, the condensation occurs within the repulsive
regime and the binding transition becomes continuous.  Thus, direct
experimental observation of the first-order binding transition may prove subtle.

\begin{figure}
\resizebox*{2.5in}{2in}{\rotatebox{0}{\includegraphics{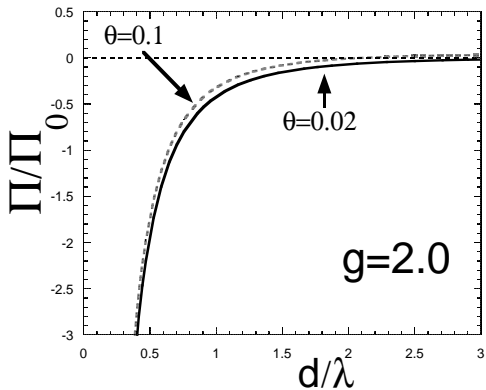}}}
\caption{The pressure profile for high surface charges.  Note that
the distance at which the pressure turns attractive can be large
compared
to the Gouy-Chapmann length $\lambda$.}
\label{largeg}
\end{figure}

Finally, For $g > g_{\infty}(\theta)$ and $\theta < \theta_c$, the
condensation again becomes continuous.  This is because the first-order phase
transition has already occurred at infinite separation, in which $\tau \sim 1$.
In this regime, the length scale at which the attraction starts to
overcome the repulsion can be quite large (see Fig. \ref{largeg}).  In the case of divalent
counterions, we find that the onset of attraction occurs at $d \sim 100$ \AA.
Clearly, the higher the surface charge or $g$, we have longer the range of the
attraction; therefore, together with the mechanism of fluctuation-driven counterion
condensation, the correlated attraction may explain the long-ranged attractions
observed experimentally. Moreover, we note that there is a qualitative change in the shape of
the pressure: the repulsive barrier disappears.  This may mark an onset of the
aggregation and has important experimental manifestations on the phase
behavior of the macroions.

In summary, by incorporating the condensation driven by fluctuations,
we show that the net pressure between two similarly charged surfaces becomes negative, hence
attractions, at a length scale much longer than the Gouy-Chapmann
length.  We also predict several distinct behaviors of the system,
depending on the valence of the counterions, that deviates significantly from the classical
theory of the double-layer interactions.  While our calculation is
based on the Gaussian fluctuation theory which may break down for very
high surface charge density, a complementary treatment is considered by
Shklovskii \cite{shklovskii} in this regime, where the condensed
counterions are assumed to form a 2D Strongly Correlated Liquid.  That theory
predicts a strongly reduced surface charge and exponentially large renormalized
Gouy-Chapmann length, qualitatively similar to our results that for high surface charge
most of the counterions are condensed.  Moreover,  it was shown in Ref. \cite{lau} that by
perturbing around the low temperature Wigner crystal ground state, the long-ranged attraction
persists to be operative, independent of the ground state.  Thus, at large distances,
we believe that our picture should capture the interaction of two similarly charged
surfaces in the regime between where PB theory is appropriate (low surface charge)
and the strong coupling limit \cite{shklovskii,netz2}.

However, there remain fundamental issues to be addressed in the future.
For example, in real systems, the charged surfaces are often
characterized by discrete surface charge distribution.  In recent studies \cite{dima},
it is shown that the counterion distribution is strongly modified if
discreteness is taken into account.  In particular, the counterions tend to be more
``localized'' near the charged surface.  It remains to be seen how this affects the
condensation picture presented in this paper; it is possible that this effect may smooth out
the first-order transition.  However, we believe that a rapid variation
of the condensation reflecting the first-order transition should remain.

\begin{acknowledgments}
We wish to acknowledge J.-F. Joanny, D.B. Lukatsky,
T.C. Lubensky, S.A. Safran, and P. Sen for important discussions.
A.W.C.L. would like to thank Prof. P.-G. de Gennes for fruitful
discussions and for a wonderful hospitality at Coll\`{e}ge de France, Paris where
most of this work is completed.  PP acknowledges support from grants
MRL-NSF-DMR-0080034 and NSF-DMR-9972246. A.W.C.L. acknowledges support
from NIH under Grant No. HL67286.
\end{acknowledgments}

\appendix

\section{Derivation of the Fluctuation Pressure}
\label{app:derivation}

In this appendix, we present a detail derivation of the
pressure arising from fluctuations.  The derivative of the fluctuation
free energy with respect to the distance $d$ is given in Eq. (\ref{expressure})
\begin{widetext}
\begin{eqnarray}
{ \partial \beta \Delta {\cal F} \over \partial d }
= { 1 \over 2 \ell_B} \int d^{3}{\bf x}\, G_{3d}({\bf x},{\bf x})\,
{\partial \over \partial d } \left [ {2 \over \lambda_D
}\,\sum_{\pm}\,\delta(z \pm d/2)
+\, \kappa^2 e^{-\varphi(z)}
\right ],
\label{expressureA}
\end{eqnarray}
where $\kappa^2 e^{-\varphi(z)\,} =
2 \alpha^2 \sec^2(\alpha z)\,\Theta(z) +
2 \left (\, |z| - d/2 + \xi\,\right )^{-2}\,\tilde{\Theta}(z)$
as defined in Eq. (\ref{density}), $\xi \equiv \lambda_R/( 1+ b^2)$,
and $G_{3d}({\bf x},{\bf x}')$ is the
Green's function defined as the inverse operator of
$\hat{{\bf K}}_{3d}$ and satisfies the equation Eq. (\ref{greenwalls}),
which in Fourier space can be written
as:\[
\left [\,- {\partial^2 \over \partial z^2 }+ q^2 + {2\over \lambda_D}
\sum_{\pm} \delta(z \pm d/2) + \kappa^2 e^{-\varphi(z)\,}
\,\right ] G_{3d}(z,z';q ) = \ell_B\,\delta(z - z').
\]
The Green's function can be solved by standard technique \cite{arfken};
first,
we note that the homogeneous solutions are given by\bb
h_{\pm}^{<}(z;q) = e^{\pm q z} \left [  1 \pm {\alpha \over q}\,\tan
(\alpha z )
\right ],
\en
for $|z| < d/2$ and\bb
h_{\pm}^{>}(z;q) = e^{\pm q |z|}\left [ 1 \mp
{ 1 \over q\,\left ( |z| -d/2 + \xi \right ) } \right ],
\en
for $|z| > d/2$.  We have two cases to consider: $|z'| < d/2$ and $|z'|
> d/2$.
In the former case, we split the space into four regions: $z < -d/2$, $
-d/2 < z < z'$,
$z' < z < d/2 $, and $ z > d/2$ and write\begin{eqnarray}
G_{-}(z,z';q) &=& A(z')\,h_{-}^{>}(z;q),
\phantom{B(z')\,h_{+}^<(z; q) + C(z')\,h_{-}^<(z;q)} \mbox{for $z <
-d/2$},\\
G_{<}(z,z';q) &=& B(z')\,h_{+}^<(z; q) + C(z')\,h_{-}^<(z;q),
\phantom{A(z')\,h_{-}^{>}(z;q)}\mbox{for $ -d/2  <  z < z'$},\\
G_{>}(z,z';q) &=& D(z')\,h_{+}^<(z; q)
+ E(z')h_{-}^<(z;q),\phantom{A(z')\,h_{-}^{>}(z;q)}\mbox{for $z'< z <
d/2$}, \\
G_{+}(z,z';q) &=& F(z')\,h_{-}^{>}(z;q),
\phantom{B(z')\,h_{+}^<(z; q) + C(z')\,h_{-}^<(z;q)} \mbox{for $z >
-d/2$}.
\end{eqnarray}
The coefficients $A(z') \ldots F(z')$ are determined by
the following boundary conditions\begin{eqnarray}
& &\,\,G_{3d}(\pm d/2,z';q) = G_{3d}(\pm d/2,z';q), \\
& &\,\,\left. {\partial_z  G_{3d}(z,z';q)} \right |_{z=\pm d/2} -
\left. {\partial_z  G_{3d}(z,z';q) } \right |_{z=\pm d/2}
= {(2 / \lambda_D)}\,G_{3d}( \pm d/2,z';q),\label{boundary}\\
& &\,\,G_{3d}(z',z';q) = G_{3d}(z',z';q), \\
& &\,\,\left. {\partial_z  G_{3d}(z,z';q)} \right |_{z=z'} -
\left. {\partial_z  G_{3d}(z,z';q)} \right |_{z=z'} = \ell_B.
\end{eqnarray}
After some algebra, we obtain specifically\begin{eqnarray}
E(z') &=& { \ell_B \over 2 q}\,{ h^<_{+}(z') + {\cal M}(q)\,h^<_{-}(z')
\over
( 1 + \alpha^2/q^2)\,[ 1 - {\cal M}^2(q) ] }, \\
D(z') &=& {\cal M}(q) E(z'), \\
{\cal M}(q) &=& { e^{-qd}\,[\,(1+ b^2)^2 +  \gamma ( 1 -b^2 - q
\lambda_R) (1
+b^2 + q \lambda_R) \,] \over  (1 +b^2)^2(1 + q \lambda_R )  +
( \gamma + q\lambda_R)(1 + b^2 + q \lambda_R)( 1 -b^2 + q \lambda_R) },
\label{M}
\end{eqnarray}
where $ \gamma \equiv \lambda_R /\lambda_D = 2\tau/( 1- \tau)$.
Therefore, the Green's function $G_{3d}({\bf x},{\bf x})$ for $|z| \leq
d/2$
is explicitly given by\begin{eqnarray}
G_{3d}^<({\bf x},{\bf x}) &=& \int {d^{2}{\bf q} \over (2 \pi)^2}\,
{ \ell_B \over 2 q}\,\left \{
{ 1 + {\cal M}^2(q) \over 1- {\cal M}^2(q) } - { \alpha^2 \over q^2 }\,
{\sec^2 (\alpha z) \over 1 + { \alpha^2 /q^2 }}\,
{ 1 + {\cal M}^2(q) \over 1- {\cal M}^2(q) } \vphantom{{ 2 {\cal M}(q) \over  1- {\cal M}^2(q) }
\left [ {\left [ 1+ \alpha^2/q^2 \,\tan^2(\alpha z) \right
]\,\cosh(2qz) +
2\alpha/q\,\tan(\alpha z)\,\sinh (2qz) \over
1+ { \alpha^2 /q^2 } } \right ]}
 \right.
\\
&+& \left.{ 2 {\cal M}(q) \over  1- {\cal M}^2(q) }
\left [ {\left [ 1+ \alpha^2/q^2 \,\tan^2(\alpha z) \right
]\,\cosh(2qz) +
2\alpha/q\,\tan(\alpha z)\,\sinh (2qz) \over
1+ { \alpha^2 /q^2 } } \right ] \right \}.
\label{greenselfwall}
\end{eqnarray}
Note that the Green's function
is symmetric with respect to $z$, as expected from the symmetry of the
problem.
Similar calculation can be done for the case $|z| \geq d/2$ and the
result is\bb
G_{3d}^>({\bf x},{\bf x}) = \int {d^{2}{\bf q} \over (2 \pi)^2}\,
{ \ell_B \over 2 q}\,\left \{1 - { 1 - {\cal
L}(q)\,e^{-2q(|z|-d/2)}\,\left [\,1 + q\,( |z| -d/2 + \xi )\,\right ]^2
\over q^2 ( |z| -d/2 + \xi)^2 }
\right \},\en
where ${\cal L}(q)$ is given by\bb
{\cal L}(q) = {e^{-qd} {\cal G}(d/2) \over [h^>_-(d/2)]^2} -
{h^>_+(d/2) \,e^{-qd} \over h^>_-(d/2)},
\label{L}
\en
and\bb
{\cal G}(d/2) \equiv  {[h^<_{+}(d/2) + {\cal M}(q)\,h^<_{-}(d/2)]
[h^<_{-}(d/2) + {\cal M}(q)\,h^<_{+}(d/2)]\over ( 1 + \alpha^2/q^2)\,[
1 - {\cal M}^2(q) ]
}= { 2 q \over \ell_B}\,G_{3d}[d/2,d/2;q].
\label{G}
\en

Returning to the expression in Eq.(\ref{expressureA}),
we note that it can be separated into three
parts:\begin{eqnarray}
{ 1\over {\cal A}} { \partial \beta \Delta {\cal F}_{A}
\over \partial d } & = &  { 2 \over \ell_B \lambda_D }\int{d^2{\bf q}
\over
(2 \pi)^2 }\,\int_0^{\infty} dz\,G_{3d}(z,z;q)\,{\partial \over
\partial d }\, \delta(z - d/2), \label{A}\\
{ 1\over {\cal A}} { \partial \beta \Delta {\cal F}_{B} \over \partial
d }
&=& { 1\over \ell_B } \int{d^2{\bf q} \over
(2 \pi)^2 }\,\int_0^{\infty}dz\, G_{3d}(z,z;q)\, \Theta(z)\,{\partial
\over \partial d }
\left [\, 2 \alpha^2 \sec^2 (\alpha z)\,\right ], \label{B}\\
{ 1\over {\cal A}} { \partial \beta \Delta {\cal F}_{C} \over \partial
d }
&=& { 1\over \ell_B } \int{d^2{\bf q} \over
(2 \pi)^2 }\,\int_0^{\infty} dz\,G_{3d}(z,z;q)\,\tilde{\Theta}(z)\,
{\partial \over \partial d }
\left [ \,{2 \over \left (\, z - d/2 + \xi\,\right )^{2}} \right ],
\label{C}
\end{eqnarray}
where we have used the fact that the integrand is symmetric with
respect to $z$.  Note
also that there should also be two terms containing
$\partial \Theta(z) / \partial d$ and $ \partial \tilde{\Theta}(z)
/\partial d$
in ${ \partial \beta \Delta {\cal F}_{B}/\partial d }$ and
${ \partial \beta \Delta {\cal F}_{C}/\partial d }$, respectively;
however, they cancel identically when they are added together.

Let us first discuss Eq. (\ref{A}); using the identity:
$ {\partial \over \partial d }\,\delta(z - d/2) \equiv
-\,{1 \over 2} {\partial \over \partial z }\,\delta(z - d/2)$,
and integrating by part, it can be transformed into\bb
{ 1\over {\cal A}} { \partial \beta \Delta {\cal F}_{A}
\over \partial d } = { 1\over \ell_B \lambda_D } \int{d^2{\bf q} \over
(2 \pi)^2 }\,\left. {\partial G_{3d}(z,z;q) \over \partial
z}\right|_{z=d/2^{+}}.
\label{pressure1}
\en
Using the boundary condition in Eq. (\ref{boundary})\[
\left. {\partial_z G_{3d}(z,z;q)}\right|_{z=d/2^{+}} =
\left. {\partial_z  G_{3d}(z,z;q)}\right|_{z=d/2^{-}}
+ {(2 / \lambda_D)}\,G_{3d}(d/2,d/2;q),
\]
and the explicit expression for the Green's function given in Eq.
(\ref{greenselfwall}), we obtain after some algebra\begin{eqnarray}
{ 1\over {\cal A}} { \partial \beta \Delta {\cal F}_{A}
\over \partial d }  &=& { 1\over \ell_B \lambda_D } \int{d^2{\bf q}
\over (2 \pi)^2 }\,{\ell_B \over 2 q } {2 \over \lambda_D}\left
\{\, {- {\cal M}^2(q) \over 1 - {\cal M}^2(q)} {q \lambda_D (1+ b^2 )^2
( b^2 + q\lambda_R + q \lambda_D )
\over (1+ b^2)^2 +  \gamma ( 1 -b^2 - q \lambda_R) (1+b^2 + q
\lambda_R)}
\vphantom{+ {q \lambda_D \over 1 -
{\cal M}^2(q) }{ 1+ (\alpha\lambda_R /2)^2  \over 1 +
(\alpha\lambda_R /2)^2  + (\gamma + q\lambda_R)(1+q\lambda_R/2)}}
\right. \nonumber \\
&+& \left.  {q \lambda_D \over 1 - {\cal M}^2(q) }{ b^2 (1+ b^2 )^2
 \over  (1 +b^2)^2 (1 + q \lambda_R )  +
( \gamma + q\lambda_R)(1 + b^2 + q \lambda_R)( 1 -b^2 + q \lambda_R)}\,
\right \} \nonumber \\
&+& \int {d^{2}{\bf q} \over (2 \pi)^2}\,{ q {\cal M}^2(q) \over 1-
{\cal M}^2(q)}.
\label{pressureA}
\end{eqnarray}
The next term, Eq. (\ref{B}), can be shown to be\bb
{ 1\over {\cal A}}{ \partial \beta \Delta {\cal F}_B \over \partial d }
= - {8 \over \ell_B\,\lambda_R^2}\,{(1+b^2) b \over 2 + { d \over
\lambda_R}( 1 + b^2) } \int {d^{2}{\bf q} \over (2 \pi)^2}\,
\,\int_0^{\alpha d/2} dx\,G_{3d}(x,x;q)\,\sec^2 x \,\left ( 1 + x \tan
x \right ).
\label{pressure2}
\en
In evaluating the $x$-integral, we note that there is a nontrivial
integral
which involves the last term inside the bracket of Green's function
in Eq. (\ref{greenselfwall}); it
reads\[
{\cal Q} = \int_0^{\tilde{d}} dx\,
\sec^2 x ( 1 + x \tan x)\,\left \{
\left [ 1 + (2/k)^2 \,\tan^2 x \right ]\,\cosh kx +
4/k\,\tan x\,\sinh k x \right \},
\]
where $k \equiv  2q/\alpha$ and $\tilde{d} \equiv \alpha d/2$.
Note that none of these integrals can be expressed
in terms of elementary functions, but integrating by parts several
times, one can show that
the integral  ${\cal Q}$ can be expressed in closed form with the
help of the relation: $ 2 b \tan (\alpha d/2) = 1 - b^2$,
by\begin{eqnarray}
{\cal Q} &=& {(1-b^2)^2 (1 + b^2)  \over 16\, b\, (q \lambda_R)^2}\,
\left [ 2 + {d \over \lambda_R }\,
\left ( 1 +  b^2 \right ) \right ]\cosh (qd) - {b (1 + b^2)  \over 4 (q
\lambda_R)^2}
\left [ 2 + {d \over \lambda_R }\,
\left ( 1 +  b^2 \right ) \right ]\cosh (qd) \nonumber \\
&-& { (1-b^2)^2 b \over 4 (q \lambda_R)^2}\,\cosh (qd)
+ {b \,\cosh (qd) \over (q \lambda_R)^2} + {(1-b^2) (1 + b^2)  \over
8\, b\, (q \lambda_R)}\,
\left [ 2 + {d \over \lambda_R }\,
\left ( 1 +  b^2 \right ) \right ]\sinh (qd)\nonumber \\
&+& {b (1 + b^2)  \over 2 (q \lambda_R)} \sinh (qd).
\nonumber
\end{eqnarray}
Substituting this result back into Eq. (\ref{pressure2}) and
rearranging terms, we obtain\begin{eqnarray}
&&{ 1\over {\cal A}}{ \partial \beta \Delta {\cal F}_B \over \partial d
}
= - {(1 + b^2)^2 \over \ell_B\,\lambda_R^2}\,
\int {d^{2}{\bf q} \over (2 \pi)^2}\,{\ell_B \over 2 q }
\left [ {{\cal G}(d/2)\over 2}\, +  {\cal J}(d/2) \right ] +
 {1 \over \ell_B\,\lambda_R^2}\,{4 b^2\,(1+b^2) \over 2 + { d \over
\lambda_R}( 1 + b^2) } \times \nonumber \\
& & \phantom{++}\int {d^{2}{\bf q} \over (2 \pi)^2}\,{\ell_B \over 2 q
}
\left [ {{\cal G}(d/2)\over 2}\, +  {\cal J}(d/2)  +
{2 (1-b^2)\,{\cal J}(d/2) \over (q \lambda_R)^2 \left (1 + \alpha^2
/q^2\right )}\,
- {2 \over q \lambda_R}{ 2 {\cal M}(q) \, \sinh (qd)\over \left (1 +
\alpha^2 /q^2 \right )
\left [ 1 - {\cal M}^2(q) \right ]} \right ]\nonumber
\end{eqnarray}
where ${\cal J}(d/2)$ is defined by the expression\bb
{\cal J}(d/2) \equiv  {1 \over 2}\,\left [ { 1+ {\cal M}^2(q) \over 1 -
{\cal M}^2(q) } -
{2 {\cal M}(q) \,\cosh qd \over 1 - {\cal M}^2(q) }\,\right ].
\en

Finally, we turn to the last term in Eq. (\ref{expressureA}), Eq.
(\ref{C}).  With the
help of the integral\bb
\int_{d/2}^{\infty} dz\,
{\,G_{3d}(z,z;q) \over \left (\, z - d/2 + \xi\,\right )^{3}}
= {\ell_B \over 2 q }{ (1 + b^2)^2 \over 2\lambda_R^2 } \left[\,{1
\over 2}\,{\cal G}(d/2)
+  {1 \over 2} -  {1 \over 2}\,{\cal L}(q)\,\right ],
\en
Eq. (\ref{C}) can be written as\begin{eqnarray}
{ 1\over {\cal A}}{ \partial \beta \Delta {\cal F}_C \over \partial d }
&=&
{ (1+b^2)^2\over \ell_B \lambda_R^2} \int{d^2{\bf q} \over
(2 \pi)^2 }\,{\ell_B \over 2 q}\left[\,{{\cal G}(d/2) +1 -  {\cal
L}(q)\over 2}\,\right ]\nonumber \\
&-&  {4 b^2\,(1+b^2)\over 2 + { d \over \lambda_R}( 1 + b^2) }
\int{d^2{\bf q} \over
(2 \pi)^2 }\,{\ell_B \over 2 q}\left[\,{{\cal G}(d/2) +1 -  {\cal
L}(q)\over 2}\,\right ],
\end{eqnarray}
which can be combined with the expression for ${ \partial \beta
\Delta {\cal F}_B /\partial d }$ above (note that ${\cal G}(d/2)$
cancels nicely)
to yield\bb{ 1\over {\cal A}}{ \partial \beta \Delta {\cal F}_{B+C}
\over \partial d }
= {(1 + b^2)^2 \over \ell_B\,\lambda_R^2} {\ell_B \over 4 \pi
\lambda_R}\,
{\cal I}_3 + {1 \over \ell_B\,\lambda_R^2}\,{4 b^2\,(1+b^2) \over 2 + {
d \over
\lambda_R}( 1 + b^2) }{\ell_B \over 4 \pi \lambda_R}\left ( {\cal I}_2
- {\cal
I}_3 \right ),
\label{combined}
\en
where we have defined the following dimensionless
integrals\begin{eqnarray}
{\cal I}_2 &\equiv& \int {d^{2}{\bf q} \over (2 \pi)^2}\,{4 \pi
\lambda_R \over 2 q }\,
\left [ {2 (1-b^2)\,{\cal J}(d/2) \over (q \lambda_R)^2 \left (1 +
\alpha^2 /q^2 \right )}\,
- {2 \over q \lambda_R}{ 2 {\cal M}(q) \, \sinh (qd)\over \left (1 +
\alpha^2 /q^2 \right )
\left [ 1 - {\cal M}^2(q) \right ]}\, \right ],\label{I2}\\
{\cal I}_3   &\equiv&  \int {d^{2}{\bf q} \over (2 \pi)^2}\,
{4 \pi \lambda_R \over 2 q}\left[ {1 \over 2} -
{1 \over 2}\,{\cal L}(q) - {\cal J}(d/2) \right ].
\label{I3}
\end{eqnarray}
With some straightforward but tedious algebra, they can be cast into
more explicit form:
\begin{eqnarray}
&&{\cal I}_2[d/\lambda_R] = \int_0^{\infty}dx\,
{2 x \left [( 1- b^2 + x)^2 + \gamma ( 1+ b^2 +x)( 1 -b^2 +x) - b^2\,
\left (\, 4b^2 +x^2\,\right) \right ]
\over \left ( 4 b^2 + x^2 \right ) \left [ 1 - {\cal M}^2(x) \right ]
\left [
(1 +b^2)^2(1 + x) + ( \gamma + x)(1 + b^2 + x)( 1 -b^2 + x) \right ]
}\nonumber \\
& &- \int_0^{\infty}dx\,{2 x {\cal M}^2(x) \left \{( 1- b^2 -x)
\left [ ( 1 -b^2 +x) + \gamma ( 1+ b^2 +x) \right ]
- \left (\,b^2 + x \right )  \,\left (\, 4b^2 +x^2\,\right) \right \}
\over \left ( 4 b^2 + x^2 \right ) \left [ 1 - {\cal M}^2(x) \right ]
\left [
(1+ b^2)^2 +  \gamma ( 1 -b^2 - x) (1+b^2 + x)\right ] },
\label{I2fin}
\end{eqnarray}
and \begin{eqnarray}
{\cal I}_3[d/\lambda_R] &=& - \int_0^{\infty}dx\,
{2 \gamma x b^2 \over \left [ 1 - {\cal M}^2(x) \right ] \left [
(1 +b^2)^2(1 + x) + ( \gamma + x)(1 + b^2 + x)( 1 -b^2 + x) \right ] }
\nonumber \\
&+& \int_0^{\infty}dx\,{2 x {\cal M}^2(x) \left [\,x + \gamma ( b^2 +x)
\right ]
\over \left [ 1 - {\cal M}^2(x) \right ] \left [ (1+ b^2)^2 +
\gamma ( 1 -b^2 - x) (1+b^2 + x)\right ] },
\label{I3fin}
\end{eqnarray}
where we have made a change of the integration variable $ x = q
\lambda_R$.
Now, observe that the first term in Eq. (\ref{combined}) cancels
precisely the first term
in Eq. (\ref{pressureA}). Therefore, combining the two expressions,
we obtain Eq. (\ref{pressurefin}) for the fluctuation pressure.
Similarly, ${\cal I}_1[d/ \lambda_R]$
defined in Eq. (\ref{I1}) can be expressed as:

\begin{eqnarray}
&& {\cal I}_1[d/ \lambda_R] = - \int{d^2{\bf q} \over
(2 \pi)^2 }\,{4 \pi \lambda_R \over 2 q }
{q \lambda_R \left [ (1+b^2)^2 + 2 \gamma (1-b^2 + q \lambda_R )\right
]
\over (1 +b^2)^2(1 + q \lambda_R )  +
( \gamma + q\lambda_R)(1 + b^2 + q \lambda_R)( 1 -b^2 + q \lambda_R) }
\nonumber \\
& & - \int{d^2{\bf q} \over
(2 \pi)^2 }\,{4 \pi \lambda_R \over 2 q }\,{  q \lambda_R {\cal
M}^{2}(q)
\over  1 -{\cal M}^2(q) } \left \{ {(1+b^2)^2 + 2 \gamma (1-b^2 + q
\lambda_R )
\over (1 +b^2)^2(1 + q \lambda_R )  + ( \gamma + q \lambda_R)(1 + b^2 +
q \lambda_R)
( 1 -b^2 + q \lambda_R)  } \right. \nonumber \\
&& \phantom{+++++++++}
 - \left. {(1+b^2)^2 + 2 q \lambda_R (1-b^2 ) + 2(q \lambda_R)^2 +
2\gamma (1-b^2 -q \lambda_R)
 \over(1 +b^2)^2  + \gamma(1 - b^2 - q \lambda_R)( 1 + b^2 + q
\lambda_R) } \right \}.
\label{I1fin}
\end{eqnarray}
\end{widetext}
Note that the first term in this expression is logarithmically
divergent.


\begin{thebibliography}{0}
\bibitem{mean}
Note that this is the case where the counterions can be squeezed out of
the surfaces, where the density at the mid-plane for $d \ll \lambda$,
$\rho_0(0) \sim 1/(l_B\,\lambda^2 )$, instead of  $\rho_0(0) \sim
1/(l_B \lambda\,d)$ where counterions are confined between two charged hard walls.
\end{thebibliography}
\end{document}